\documentclass[preprint2]{aastex}

\usepackage{amsmath}
\usepackage{amssymb}
\usepackage{bm}

\shorttitle{Electron-Positron Reconnection}
\shortauthors{Swisdak, Liu, and Drake}

\begin{document}

\title{Development of a Turbulent Outflow During Electron-Positron
Magnetic Reconnection}

\author{M. Swisdak, Y. H. Liu, and J. F. Drake\altaffilmark{1}}
\affil{IREAP, University of Maryland,
    College Park, MD 20740-3511}
\email{swisdak@umd.edu}

\altaffiltext{1}{SSL, University of California, Berkeley, CA 94720-7450}

\begin{abstract}
The mass symmetry between the two species in electron-positron (pair)
plasmas has interesting consequences for collisionless magnetic
reconnection because the Hall term, which plays a crucial role in
supporting fast reconnection in electron-proton plasmas, vanishes.  We
perform kinetic simulations of pair reconnection in systems of various
sizes, show that it remains fast, and identify the reason why this
occurs. For sufficiently large systems a Weibel-like temperature
anisotropy instability develops in the outflow from the X-point that
causes the current layer to broaden and form a Petschek-like open
outflow. We discuss the parameter regimes in which pair reconnection
should be fast and the implications for astrophysical pair plasmas.
\end{abstract}


\section{Introduction}

The energy density of magnetic fields can be significant, and even
dominant, in certain astrophysical systems and so the question as to
which processes act as sinks and convert that energy into other forms
naturally arises.  Observations (including in situ measurements) show
that magnetic reconnection, in which topological changes in an
embedded field lead to increases in a plasma's kinetic and thermal
energy, plays such a role in the terrestrial magnetosphere, the solar
wind, and the solar corona. \citep{mozer02a,phan06a,masuda94a} The
natural extension to other astrophysical locations has led to the
consideration of the role of reconnection in accretion disks
\citep{eardley75a,miller97a}, gamma-ray bursts \citep{drenkhahn02a}
and pulsar winds \citep{coroniti90a,michel94a,zenitani01a}. Unlike the
heliosphere, positrons, rather than protons, can comprise the
positively charged species in the plasmas of these systems.  The
equality of the electron and positron masses introduces a symmetry
that can have a potentially large effect on the development and role
of reconnection.

Parker and Sweet \citep{parker57b,sweet58a} proposed the first model
of electron-proton reconnection, a magnetohydrodynamic (MHD)
formulation that attempted to explain the fast timescales associated
with solar flares.  However contemporaneous observations suggested
that magnetic flux reconnected much faster than the model predicted
(on timescales of minutes, not weeks).  Relaxing the assumptions of
classical resistive MHD by invoking an anomalously large and localized
resistivity, perhaps driven by turbulence, resolved the theoretical
discrepancy, but no model of the process garnered widespread support.

Later work focused on Hall MHD, an extension of MHD that includes the
decoupling of the motions of protons and electrons at small spatial
scales --- on the order of the ion skin depth, $d_i \equiv
c/\omega_{pi}$, where $\omega_{pi}$ is the proton plasma frequency.
The GEM Challenge, a collaboration within the reconnection simulation
community, showed that various algorithms including the Hall term (the
simplest new term arising in Hall MHD) all reconnected at
approximately the same rate: $v_{\text{in}} \approx 0.1 v_A$, where
$v_{\text{in}}$ is the speed at which plasma flows into the X-point
and $v_A$ is the local Alfv\'en speed \citep{birn01a}.  A resistive
MHD code simulating the same problem reconnected flux at a rate
several orders of magnitude slower.

Although the GEM Challenge demonstrated that inclusion of the Hall
term is a sufficient requirement for fast reconnection, the question
of whether it is a necessary requirement remains open.  Pair
reconnection provides an excellent test because one effect of the mass
symmetry is the disappearance of the Hall term from the governing
equations.  If it is necessary for fast reconnection then pair plasmas
will reconnect slowly, on Sweet-Parker timescales, and magnetic
reconnection will be an inefficient mechanism for releasing the
magnetic energy in pair plasmas.

Despite this interesting limit the simulation community has only
recently begun to consider pair reconnection.
\cite{bessho05a,bessho07a} observed fast pair reconnection in a small
system and attributed it to a localized resistivity-like effect caused
by the off-diagonal components of the pressure tensor.  
\cite{daughton07a} did a careful examination of pair reconnection in
large systems and arrived at similar conclusions for the early,
impulsive phase of reconnection.  Over longer periods they suggested
that generation of large numbers of magnetic islands served to
localize the current layer and support fast reconnection.
\cite{zenitani01a}, \cite{jaroschek04a}, and \cite{fujimoto06a}
explored other aspects of pair reconnection but, although they
observed fast reconnection, did not advance any causal explanations.

We also find that fast reconnection occurs in pair plasmas.  However,
we reach different conclusions from \cite{daughton07a} on the role of
magnetic islands.  In the initial phase of electron-positron
reconnection many magnetic islands typically form, especially if the
ratio of the initial width of the current layer to the length of the
system is very small. These islands have a transient influence on the
development of reconnection until they convect downstream. Secondary
islands also form later in time, and in greater numbers than in 
electron-proton reconnection. But, we find that these islands do
not seem to control the overall length of the current layer and
therefore the rate of reconnection.  Based on our simulations we
instead propose that the outflow jet in large pair plasma systems
opens as a result of turbulence.  We show that as the system size
grows the current layer lengthens, just as Hall-less Sweet-Parker
theory predicts.  Interestingly the reconnection rate remains constant
during this change for small systems because other parameters,
primarily the speed of the outflow jet, also vary.  This
cannot continue for arbitrarily large boxes because of the Alfv\'enic
limitation on the outflow speed.  Instead, once the system is big
enough (and the current layer long enough) an electromagnetic
pressure-anisotropy instability similar to the Weibel mode begins to
grow in the outflow jet.  The outflow jet becomes strongly turbulent
and the outflow layer broadens downstream as proposed by Petschek
\citep{petschek64a}. We present a simple analytic argument that
suggests that as a result of this turbulence the rate of reconnection
should take on a constant value, independent of system size.

In Section \ref{comps} of this paper we describe the details of our
computations.  Section \ref{hall} outlines the role of the Hall term
and how it disappears in pair plasmas.  We discuss the simulations in
Section \ref{sims} and the instability and its consequences in Section
\ref{instability}.  Finally, Section \ref{discussion} explores some
implications of our findings.

\section{Computational Details}\label{comps}

We use p3d, a massively parallel particle-in-cell (PIC) code
\citep{zeiler02a} to perform our simulations.  As in other PIC codes we
divide the computational domain into cells, track the electromagnetic
fields only on the gridpoints, and allow the particles to move freely.
During a timestep the Boris algorithm advances the Lorentz equation of
motion for each particle: the electric field $\mathbf{E}$ provides an
acceleration for half a timestep, the magnetic field $\mathbf{B}$
rotates the velocity vector, and $\mathbf{E}$ accelerates for the
second half-timestep.  An explicit trapezoidal-leapfrog method
employing second-order spatial derivatives in Faraday's and Amp\`ere's
Laws advances the fields in time.  We use uniform, square
computational cells that force $\bm{\nabla \cdot}
\mathbf{B} = 0$.  However our discretization scheme leads to a
discrepancy between $\bm{\nabla \cdot} \mathbf{E}$ and $4\pi \rho$.
To address this problem we use a multigrid algorithm to solve
Poisson's equation for a correction term to $\mathbf{E}$.  Our
particles do not directly interact and, in particular, no
electron-positron annihilation occurs.

To help elucidate the underlying physics we write the code equations
in normalized units.  Masses are normalized to the ion mass $m_i$,
magnetic fields to the asymptotic value of the reversed field, and the
density to the value at the center of the current sheet.  Other
normalizations derive from these: velocities to the Alfv\'en speed
$v_A$, lengths to the ion inertial length $c/\omega_{pi}$, times to
the inverse ion cyclotron frequency $\Omega_{ci}^{-1}$, and
temperatures to $m_i v_A^2$.  In electron-positron plasmas $m_e/m_i =
1$.

In our coordinate system the inflow and outflow for an X-point lie
parallel to $\mathbf{\hat{y}}$ and $\mathbf{\hat{x}}$, respectively.
The reconnection electric field is parallel to $\mathbf{\hat{z}}$.  In
the simulations presented here we assume out-of-plane derivatives
vanish, i.e., $\partial/\partial z = 0$; this choice eliminates any
structure in the $\mathbf{\hat{z}}$ direction.  The reasonableness of
simulating pair plasmas with this restriction remains unknown.  For
electron-proton plasmas, a comparison of 2D and 3D reconnection
simulations suggests that the qualitative features remain unchanged
\citep{hesse01a}. However for many instabilities that can have large
effects on the out-of-plane structure, e.g., the drift-kink, the onset
threshold varies with the mass ratio, and so further simulations are
needed to explore their importance in the $m_e/m_i=1$ limit.

The initial equilibrium consists of two Harris current sheets
\citep{harris62a} superimposed on a ambient population of uniform
density.  The reconnecting magnetic field is given by
$B_x/B_0=\tanh[(y-L_y/4)/w_0]- \tanh[(y-3L_y/4)/w_0]-1$, where $w_0$ and
$L_y$ denote the half-width of the current sheets and the box size in
the $\mathbf{\hat{y}}$ direction respectively.  This configuration
allows us to use fully periodic boundary conditions.  Both species
have the same uniform initial temperature, $T_e = T_i =
0.25$. Pressure balance uniquely determines the density profile, aside
from a uniform background population that can have arbitrary
density (here $n_b=0.2$); in this equilibrium $n(y=L_y/4) =
n(y=3L_y/4) = 1.2$.  At $t=0$ we perturb the magnetic field ---
$\widetilde{B}_x/B_0$ varies with system size but is always $< 0.07$
--- to seed X-points at $(x,y) = (L_x/4,3L_y/4)$ and $(3L_x/4,L_y/4)$.

We set the speed of light (in normalized units) to $5$.  The spatial
resolution is such that there are $>4$ gridpoints per inertial length
and $\approx\negmedspace1$ per Debye length.  The Courant
condition determines the particle timestep; we substep the advancement
of the electromagnetic fields.  A typical cell contains
$\sim\negmedspace200$ particles and our largest simulation follows
$>\negmedspace10^9$ particles.  All of our simulations conserve energy
to better than $1$ part in $200$.

\section{The Hall Term}\label{hall}

The current theory of electron-proton reconnection emphasizes the role
of the Hall term in the generalized Ohm's Law.  By combining the
collisionless fluid momentum equations for electrons and a singly
charged positive ion species of mass $m_i$ we can write
\begin{equation}\label{gen}
\begin{split}
\left(1+\mu\right)\mathbf{E} =
& -\frac{1+\mu}{c}\,\mathbf{v \times B} \\ & +\frac{1-\mu}{nec}\,\mathbf{J
\times B} \\ &
-\frac{1}{ne}\,\boldsymbol{\nabla\cdot}\left(\mathrm{P}_e-\mu\mathrm{P}_i\right)
\\ &+\frac{m_e}{ne^2}\left[\frac{\partial \mathbf{J}}{\partial t} +
\boldsymbol{\nabla\cdot}
\left(\mathbf{Jv}+\mathbf{vJ}-\frac{1}{ne}\frac{1-\mu}{1+\mu}\,\mathbf{JJ}\right)\right]
\end{split}
\end{equation}
where $\mu = m_e/m_i$, $\mathbf{v} =
(m_e\mathbf{v_e}+m_i\mathbf{v_i})/(m_e+m_i)$ is the velocity in the
center of mass frame, $\mathrm{P}$ is the pressure tensor, and we
assume quasi-neutrality, $n_i=n_e=n$.  For electron-proton plasmas
$\mu \approx 0$.

By scaling equation \ref{gen} one can show that the first term on the
right-hand side dominates at large lengthscales.  The MHD model, and
hence Sweet-Parker theory, discards the remaining terms.  The second
term on the right-hand side is the Hall term; it becomes important for
lengthscales equal to or smaller than $d_i$.  With the addition of the
Hall term the principal wave mode on these scales changes from the
Alfv\'en wave of MHD ($\omega \propto k$) to the dispersive ($\omega
\propto k^2$) whistler wave.  The localization of the current layer
along the outflow direction due to the action of dispersive waves has
been proposed as the mechanism that facilitates fast reconnection in
electron-proton plasmas \citep{birn01a,rogers01a}. The terms
proportional to the diagonal portion of the pressure tensor can be
important for systems with large guide fields (since they are the
source of kinetic Alfv\'en waves), while the off-diagonal terms and
final inertial term matter only on electron scales, $d_e =
c/\omega_{pe}$, and likely play only a small role outside the
immediate vicinity of the X-point and magnetic separatrices.

In pair plasmas $\mu =1$ and equation \ref{gen} simplifies to the form
derived in \cite{bessho05a}:
\begin{equation}\label{pairgen}
\begin{split}
\mathbf{E} =
& -\frac{1}{c}\,\mathbf{v \times B}  \\
& -\frac{1}{2ne}\,\boldsymbol{\nabla\cdot}\left(\mathrm{P}_e-\mathrm{P}_i\right) \\
&+\frac{m_e}{2ne^2}\left[\frac{\partial \mathbf{J}}{\partial t} +
\boldsymbol{\nabla\cdot} \left(\mathbf{Jv}+\mathbf{vJ}\right)\right]
\end{split}
\end{equation}
The mass symmetry eliminates the Hall term, and with it the usual
dispersive modes, including the whistler.  If we make the reasonable
assumption that $\mathbf{v}$ vanishes at a steady-state X-point,
equation \ref{pairgen} becomes
\begin{equation}\label{XOhm}
\mathbf{E}  = - \frac{1}{2ne}\,\boldsymbol{\nabla\cdot} (\mathrm{P}_e-\mathrm{P}_i)
\end{equation}
Hence a non-zero reconnection electric field ($E_z$) requires an
asymmetry between the two species, $\mathrm{P}_e\neq \mathrm{P}_i$.
In our quasi-two-dimensional system the balance must come from the
off-diagonal components of the pressure tensors, which describe the
transport of momentum away from the X-point.

%For what it's worth the equation of motion is given by
%%\begin{equation}\label{eom}
%%\begin{split}
%\begin{multline}
%(1+\mu)\frac{\partial (n\mathbf{v})}{\partial t} + \nabla\cdot
%\left(n(1+\mu)\mathbf{vv}+\frac{1}{ne^2}\frac{\mu}{(1+\mu)}\mathbf{JJ}\right)
%\\
%= \frac{1}{m_ic}\mathbf{J\times B}  -
%\frac{1}{m_i}\nabla\cdot(\mathrm{P}_e+\mathrm{P}_i)
%\end{multline}
%%\end{split}
%%\end{equation}

\section{Simulations}\label{sims}

\subsection{Overview}

\begin{figure}
\plotone{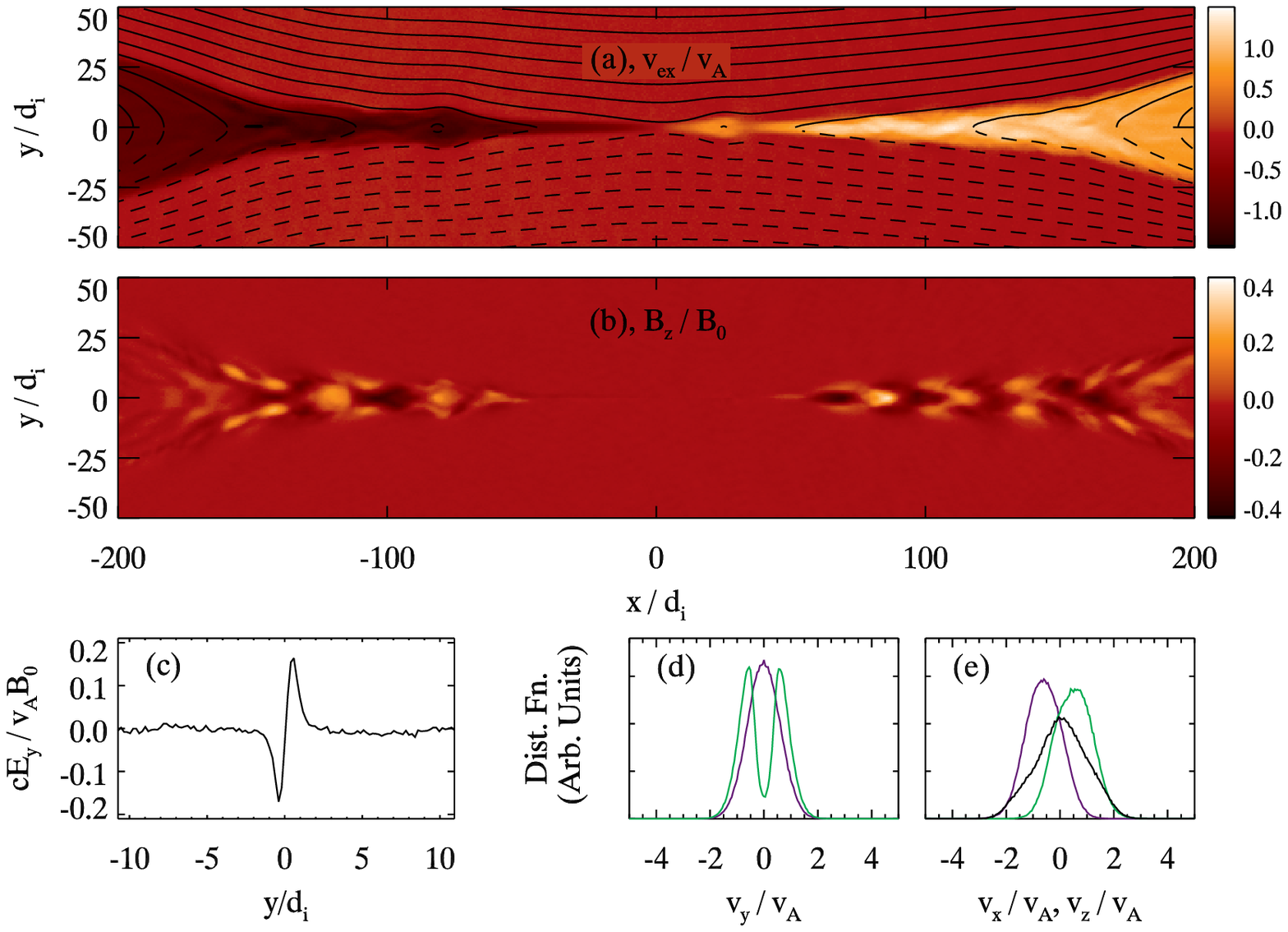}
\caption{Reconnection overview.  Panel (a): $v_{ex}$ overplotted with
magnetic field lines. The solid and dashed portions of the lines
indicate the direction of the reconnecting magnetic field $B_x$. Panel
(b): $B_z$ at the same time.  Panel (c): Cut of $E_y$ averaged
horizontally over $-50 < x < 50$.  Panel (d): Distribution of $v_y$
over the region $-50<x<50, -2<y<2$.  Electrons are green,
positrons blue.  Panel (e): $v_z$ and $v_x$ distributions.  $v_x$ is
plotted in black and is essentially the same for both
species.\label{synopsis}}
\end{figure}

Figure \ref{synopsis} shows the steady-state behavior of a typical
simulation of pair reconnection in a large domain.  The top panel
displays magnetic field lines superimposed on the component of the
electron velocity in the $\mathbf{\hat{x}}$ direction; the positron
velocity (not shown) is basically identical.  The solid and dashed
portions of the field lines indicate positive and negative signs of
$B_x$, the reconnecting component of the field.  The tension in the
highly bent field lines to the left and right of the X-point ($x=0$,
$y=0$) accelerates the plasma downstream until it reaches the local
Alfv\'en velocity ($\approx 1$ in our normalization).  A more gradual
inflow with speed $\approx 0.1$ (not shown) replenishes the plasma.

Panel (b) shows the out-of-plane component of the magnetic field
$B_z$. The quadrupolar $B_z$ pattern that is one of the signatures of
Hall reconnection is not present.  Instead $B_z$ becomes large and
highly structured farther downstream from the X-point where the current
layer opens up.  We discuss the turbulence, the instability
responsible for it, and its linkage to the structure of the outflow
jet later.

Although $E_y$ has a structure similar to $B_z$ in the turbulent
outflow, along the thin central portion of the current layer it
exhibits a narrow bipolar feature.  Panel (c) shows a vertical cut of
$E_y$ averaged over the region $-50<x<50$.  The central structure has
a width of $\approx 2 d_e$ and breaks the symmetry between the
electrons and positrons.  It grows from the ambient noise in all of
our simulations, regardless of box size, but the sign (whether
electrons or positrons accelerate toward the current layer) appears to
arise spontaneously. In this case electrons are accelerated toward,
and positrons repelled from, the center, which leads to an asymmetry
in the respective particle distribution functions.  The bipolar signal
appears much earlier in the simulations than the temperature
anisotropy instability we discuss later, and it appears in runs for
which this other instability is absent.  Instead, the
counter-streaming inflowing plasma likely triggers a two-stream
instability at the current layer in which the polarity of $E_y$ is
determined randomly.

To maintain this field the species' number densities exhibit a slight
asymmetry (i.e., quasi-neutrality is slightly violated). Panel (d)
displays the distribution of $v_y$ for the electrons (green) and
positrons (blue) within the box $|x|<50$, $|y| < 2$ and shows the
effects of this field.  The $v_z$ distributions in panel (e) are
mirror-symmetric due to the effect of $E_z$ (the reconnection electric
field) on the oppositely charged electrons and positrons.  The $v_x$
distributions are the same for both species because $E_x$ is small.
They are plotted in black.

In some of our preliminary simulations the long initial current layers
produced a larger number of islands that played a transient role
before being swept away. We find that varying the thickness of the
initial layer affects the initial rate of island production, with a
broader layer producing fewer islands.  At the time of Figure
\ref{synopsis} these initial islands have all disappeared downstream.
However, we also find that secondary islands self-generated on the
current layer remain a common occurrence in pair reconnection and
are much more prevalent than in reconnection with $m_e \ll m_i$
\citep{shay07a}.  Two such islands (at $x\approx 20$ and $x\approx
-80$) appear in Figure \ref{synopsis}. These remnant islands, however,
are not responsible for the turbulent fluctuations in $B_z$ in Figure
\ref{synopsis}. The island at $x\approx 20$, occurs in a region with
no turbulence and has no $B_z$ signature while the other, at $x\approx
-80$, lies in the turbulent outflow where any independent $B_z$
signature (if it exists) is difficult to distinguish.

\subsection{Scaling of the Reconnection Rate}

Fast reconnection ultimately arises from the localization of the
current layer \citep{biskamp01a}.  This can be seen from a
straightforward analysis of the fluid continuity equation
\begin{equation}
\frac{\partial n}{\partial t} + \boldsymbol{\nabla \cdot} (n\mathbf{v}) = 0
\end{equation}
In a steady state the time derivative vanishes, leaving a term that
the divergence theorem transforms into a surface integral.  For one
quadrant of a rectangular domain with length (parallel to
$\mathbf{\hat{x}}$) $2\Delta$ and width (parallel to
$\mathbf{\hat{y}}$) $2\delta$ centered on the X-point, the
surface integrals imply that
\begin{equation}\label{cont}
v_{\text{in}} = v_{\text{out}} \left(\frac{\delta}{\Delta}\right)
\left(\frac{n_{\text{out}}}{n_{\text{in}}}\right)
\end{equation}
where the subscripts ``in'' and ``out'' refer to inflowing and
outflowing plasmas, respectively.  Both the outflow speed, which
approaches the local Alfv\'en speed, and the density ratio
$n_{\text{out}}/n_{\text{in}}$, which does not radically change from
its initial value, are $\mathcal{O}(1)$. If, as seems a reasonable
assumption, the inflow scale $\delta$ equals an intrinsic kinetic
scale-length such as $d_e$ and remains roughly constant as the system
size changes then $v_{\text{in}}$ scales inversely with the outflow
dimension $\Delta$.  In particular, if, as in the Sweet-Parker
picture, $\Delta$ scales with the size of the system, $v_{\text{in}}$
should decrease for larger and larger simulation domains.  In
electron-proton reconnection the Hall term keeps $\Delta$ fixed,
regardless of the system size, and maintains a fast reconnection rate
$v_{\text{in}}/v_{\text{out}} \sim \delta/\Delta \sim
\mathcal{O}(0.1)$ \citep{shay99a,shay07a}.

If the Hall term is the only mechanism that can localize the current
layer then pair reconnection should feature current sheets that scale
with the system size.  In order to test this behavior we began with a
relatively small system and then doubled the size repeatedly.  Table
\ref{tbl-1} contains the parameters of the four runs we discuss in
this paper; Figure \ref{synopsis} comes from run $d$.  Because of
computational constraints we doubled $L_x$ but not $L_y$ between $c$
and $d$, and so the latter's aspect ratio differs from the other three
runs.  Earlier work showed that if the system is large enough to reach
a steady state the reconnection rate in a double tearing mode
configuration is independent of the aspect ratio \citep{shay99a}.

\begin{figure}
\plotone{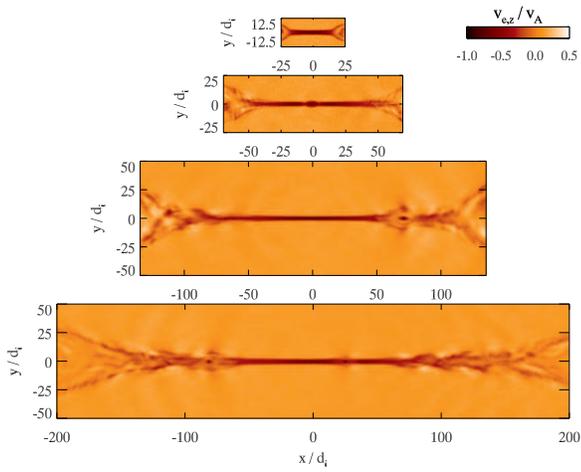}
\caption{Out-of-plane electron velocities for the simulations of Table
\ref{tbl-1}, showing the evolution of the current layer with system
size.  Each panel shows about one-fourth of the total simulation
domain.\label{compare}}
\end{figure}

Figure \ref{compare} shows the steady state current layers for the
four runs.  In the smallest simulations (panels (a) and (b)) the
current layers lengths approach the system size.  But, when the system
size doubles between panels (b) and (c) the current layer does not; it
lengthens moderately, but the turbulence evident at both ends of the
layer broadens it along the $y$ direction and prevents the narrow
portion of the layer from expanding further.  A final lengthwise
doubling of the system, panel (d), barely increases the length of the
current layer.

In Figure \ref{rec_rate} we show the reconnection rates as a function
of time.  To measure this rate we first integrate the magnetic flux
between the initial X-point and the large magnetic island formed by
the reconnected field.  (The large island is a consequence of our
periodic boundaries and is not shown in the figures.)  We check for
the growth of other X-points, but find that the initially seeded one
remains dominant throughout the runs, despite the formation of
secondary magnetic islands.  Figure \ref{rec_rate} shows the
(unsmoothed) temporal derivative of the reconnected flux. The vertical
lines mark the times of the snapshots of $v_{ez}$ displayed in Figure
\ref{compare}.  All of the runs exhibit essentially the same behavior:
after an initial quiescent period the reconnection rate rises to an
asymptotic value of $\approx 0.1$ and remains there.  Given enough
time the separatrices bounding the magnetic island on the simulation's
other current layer approach the X-points shown in Figure
\ref{compare} and we halt the run.  The first effects of this
interaction can be seen in the slight increases in the reconnection
rates noticeable at the end of each panel in Figure \ref{rec_rate}.

So, despite the lack of a Hall term, pair reconnection remains fast
when the system size increases.  A more quantitative description of
the reconnection can be found in Table \ref{tbl-2}, which lists the
various quantities from equation \ref{cont} for each run.  The
$v_{\text{in,sim}}$ column contains the inflow speed measured in the
simulation while $v_{\text{in,calc}}$ is the value calculated from the
other data using equation \ref{cont}.

For our smallest domain, run $a$, $2\Delta \approx 35$ approaches
$L_x/2 = 50$.  (The periodic boundaries in our simulations restrict
the maximum current sheet length to $L_x/2$.  In practice $\Delta$
never reaches this maximum because the transition from outflow to
island has a finite scale.)  Even though the current layer stretches
to the system size in a Sweet-Parker-like manner, reconnection remains
fast ($v_{\text{in}} \sim \mathcal{O}(0.1)$) because the box is relatively
small.  In fact the outflow velocity does not even reach the Alfv\'en
speed. Thus, exploring the scaling of reconnection in pair plasmas
requires much larger simulation domains than in normal
reconnection. In a simulation domain of this size (measured in ion
inertial lengths) the ion outflow velocity would have reached the full
Alfv\'en speed.

After doubling the box size (run $b$), the current layer length
$\Delta$ also increases by a factor of two.  However the larger box
contains more room for the acceleration of the outflow and so
$v_{\text{out}}$ also increases.  The densities, both upstream and
downstream, as well as the thickness of the current layer remain
relatively constant between these two runs, and all of our other runs.
The net effect of the variation of $\Delta$ and $v_{\text{out}}$ is to
leave the reconnection rate unchanged.  Since the outflow speed has an
upper limit, the Alfv\'en speed, this progression cannot continue
indefinitely.

After the next doubling (run $c$) the outflow velocity reaches its
expected Alfv\'enic limit.  Since the increase was less than a factor
of 2 ($1.3/0.8 \approx 1.6$) a Sweet-Parker-like doubling of $\Delta$
would be accompanied by a decrease in the reconnection rate.  Instead
of doubling, however, $\Delta$ only increases by a factor of $120/80
\approx 1.5$ and the reconnection rate stays approximately constant.
Since the Hall term cannot halt this increase, some other process must
be at work.  

As a check we again doubled the length of the box (but not its width)
for run $d$.  Unlike in the previous doublings the outflow speed
cannot increase further.  Its maximum value is the Alfv\'en speed
based on the downstream density and the upstream magnetic field; in
these simulations $\max(v_{\text{out}}) \approx 1.3$.  Absent any
other localizing effect the current layer length should approximately
double and the reconnection rate should halve.  Instead the current
layer length and the reconnection rate remain basically unchanged.  In
Section \ref{instability} we argue that the length of the current
sheet is limited by the development of an instability driven by a
temperature anisotropy.  The instability develops in the outflow from
the X-point before eventually growing strong enough to open the layer.

\section{Anisotropy Instability}\label{instability}

In sufficiently large simulations the outflow from the X-point is
subject to a temperature anisotropy instability.  The top panel of
Figure \ref{texx} shows a view $B_z$ from run $c$ at $t = 450$, soon
after the instability develops.  In the second panel we plot two of
the diagonal components of the positron temperature tensor, $T_{xx}$
and $T_{yy}$, along the line $y=-1$.  (The influence of the $E_y$
feature shown in Figure \ref{synopsis} makes the anisotropy somewhat
weaker exactly on the symmetry axis.)  At the X-point ($x=0$) the
plasma temperature is isotropic and equal to the initial value of
$0.25$.  As the plasma moves away from the X-point $T_{xx}$ increases
sharply while $T_{yy}$ remains nearly unchanged, leading to an
anisotropy that reaches a peak value of $\approx\negmedspace 4$.  At
$|x| \approx 50$ the disturbance in $B_z$ becomes visible, indicating
onset of an instability.  Further downstream the $B_z$ pattern
strengthens and the anisotropy simultaneously weakens. Movies of $B_z$
(not shown) demonstrate that the disturbances propagate with the mean
flow of the plasma in the outflow jet.

In order to investigate the source of the anisotropy we show the 1-D
velocity distributions taken between the dashed lines in panel (b) and
$-2<y<-1$ in the final three panels.  Two populations intermingle in
this region, the inflowing plasma with $v_x \approx 0$ and a jet with
$v_x < 0$ moving outward from the X-point.  As can be seen in panel
(c) the superposition of these populations results in a wide, hot
distribution of $v_x$.  Meanwhile, since the sample box is upstream of
the bipolar $E_y$ signal, the $v_y$ distributions remain essentially
unchanged from their initial Maxwellians.  A mixture of the inflowing
ambient plasma and plasma accelerated by the reconnection electric
field form the $v_z$ distributions.  The final result is a plasma with
$T_x > T_y \approx T_z$.

Plasmas with hot bimaxwellian distribution functions are typically
stable to electrostatic perturbations (the thermal spread stabilizes
the electrostatic ion-ion streaming instability), but unstable to
electromagnetic instabilities. One possibility is that the magnetic
perturbations in Figures \ref{synopsis} and \ref{texx} result from the
firehose instability. If that were the case the perturbed motion of
the plasma in the $z$ direction with ${\bf k}=k_x\hat{{\bf x}}$ would
produce perturbations in $B_z$. We can rule out this instability,
however, since along the symmetry axis ($y=0$) motion of the plasma in
$z$ would not produce a perturbation of $B_z$ since the equilibrium
field $B_x$ is zero in this region. The simulations clearly have large
perturbations of $B_z$ centered at $y=0$. Instead the anisotropy is
sufficiently large that the magnetic field in the region around $y=0$
is negligible and the instability is a Weibel-like mode that self
generates the magnetic field perturbation by separating the electrons
and positrons moving in opposite direction to create a current in the
$x$-direction that varies with $y$. To demonstrate the viability of
the instability we begin with a tri-maxwellian plasma described by the
distribution function
\begin{equation}
f = \prod_{\alpha = {x,y,z}}\left(\frac{m}{2\pi
k_BT_{\alpha}}\right)^{1/2}
\exp\left[-\left(\frac{m}{2\pi k_B T_{\alpha}}\right) v_{\alpha}^2 \right]
\end{equation}
with, in general, $T_x \neq T_y \neq T_z$, and look for instabilities
with $\widetilde{B}_z$ and $\widetilde{E}_x$ varying as $\exp(ik_y y - i\omega
t)$. Following the method of \cite{krall86b}, the instability
condition for a pair plasma is
\begin{equation}\label{growthcondtion}
\left(\frac{T_x}{T_y}-1\right) > \frac{k_y^2c^2}{2\omega_{p}^2}
\end{equation}
Setting the plasma dielectric function to zero, we obtain the
dispersion relation relating the growth rate $\gamma = -i\omega$ and
the wavenumber:
\begin{equation}\label{disp}
k_yc^2 + \gamma^2 + 2\omega_{p}^2\left(1-\frac{T_x}{T_y}\right) =
2\omega_p^2\frac{T_x}{T_y}\,\zeta\,Z(\zeta)
\end{equation}
where the thermal speed $v_{\text{th}}$ enters through
\begin{equation}
\zeta = \frac{\omega}{k_y\sqrt{2T_y/m}}\equiv \frac{\omega}{k_yv_{\text{th}}}
\end{equation}
and $Z(\zeta)$ is the plasma dispersion function
\begin{equation}
Z(\zeta) =
\frac{1}{\sqrt{\pi}}\int^{\infty}_{-\infty}\frac{\exp(-t^2)}{t-\zeta}\,dt
\end{equation}
In the limit $|\zeta| \gg 1$ equation \ref{disp} has a solution for the
growth rate of
\begin{equation}\label{limit}
\gamma \approx
k_yv_{\text{th},x}\frac{\omega_p}{\sqrt{k_y^2c^2+2\omega_p^2}}
\end{equation}
where $v_{\text{th},x}=\sqrt{2T_x/m}$. Thus, the growth rate
asymptotes to $\gamma \sim v_{\text{th},x}/d_e$ for
$k_yd_e>\sqrt{2}$. Taking $|\zeta| \gg 1$ is only possible if the
anisotropy is strong, $T_x/T_y \gg 1$. Equation
\ref{limit} can also be derived in the fluid limit by considering the
stability of a cold plasma beam.

\begin{figure}
\plotone{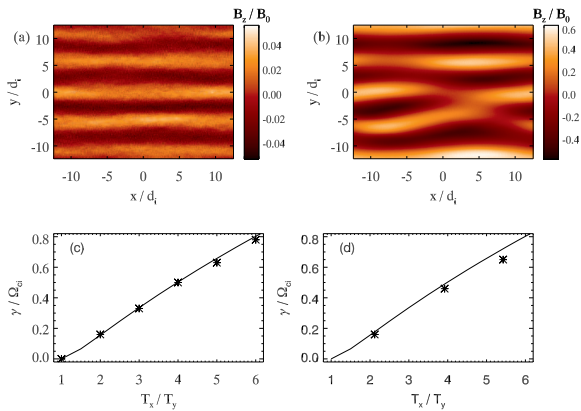}
\caption{Panels (a)-(b): The structure of $B_z$ during the linear phase
of the instability and after saturation in an initially
homogeneous plasma with $T_{x}/T_{y} = 4$.  Panel (c): Theoretical
(line) and measured (star) growth rate versus temperature anisotropy
for the fastest growing instability in a homogeneous plasma (finite
$k_y$ with $k_x=0$).  Panel (d): Growth rate in a Harris-sheet
background.  Stars indicate the measured values, the line the
theoretical value in a homogeneous medium.\label{inst}}
\end{figure}

Strictly speaking these results apply only to the case of a
homogeneous infinite plasma, while in the case under consideration
here the instability develops within a thin (width $2\delta \approx 5
d_e$) current layer.  To further explore the behavior of this
instability, particularly the behavior of its growth rate, we again
turn to simulations.  First, in order to test our code's ability to
reproduce the results of equation \ref{disp}, we consider a
homogeneous electron-positron system with a bimaxwellian distribution
(for both species), no initial magnetic field, and a constant particle
density, $n=1.2$.  We seed the system with the theoretical fastest
growing wavenumber (with $k_x=0$) and monitor the growth of the
instability.  In panel (c) of Figure \ref{inst} we plot the growth
rates calculated from the simulations (stars) compared with the values
predicted by equation \ref{disp} (solid curve). The two are in near
perfect agreement.

In order to study the effect of the narrow current layer on the
development of the instability we next turn to a more complicated
system.  We begin with a Harris sheet equilibrium similar to that
discussed in Section \ref{comps} except that one component of the
temperature, $T_{x}$, increases along with the density in the center
of the current layer. Note that $T_{x}$ does not affect force balance
in the $y$ direction and therefore does not affect the initial
equilibrium. For our usual box sizes this equilibrium is also unstable
to the tearing mode, so we consider narrow domains (e.g., $L_x = 4$,
$L_y = 50$) where the tearing mode is stabilized but the temperature
anisotropy mode is not.  In panel (d) of Figure \ref{inst} we show the
growth rates (marked by the stars) for the instability in this
geometry (again with $k_x=0$) compared with the homogeneous growth
rate (solid curve).  In general the growth rates are only reduced
around $ 20\,\%$ from the homogeneous values. Thus, neither the finite
geometry of the current layer nor the finite magnetic field in the
layer significantly affect the growth of the instability, confirming
that the Weibel instability can exist in a magnetically supported
current layer.

In panels (a) and (b) of Figure \ref{inst} we show the nonlinear
development of the instability in a 2-D homogeneous system with no
initial magnetic field and $T_{x}/T_{y}=4$. The system is seeded with
a magnetic perturbation corresponding to the fastest growing mode that
dominates the structure at early time. At late time the dominant mode
develops a finite value of $k_x$ ($k_x=0.25$, $k_y=1.0$), as also
occurs in the reconnection simulations. The amplitude of the late-time
perturbations of $B_z$ is about $0.5$; compared with the turbulence
seen in the reconnection simulations these perturbations are at
somewhat longer wavelength and have modestly larger amplitudes. These
differences may reflect constraints on the turbulence due to the
finite width (in the $y$ direction) of the temperature anisotropy in
the reconnection simulations.

For the full reconnection simulation a typical temperature anisotropy
in the outflow is $\approx 3$ (see Figure \ref{texx}), which
corresponds to a typical growth rate of $\gamma \approx 0.2$. (The
density in the reconnection outflow is $\approx 4$ times smaller than
in the instability simulations, which, since $\gamma \propto
\omega_p$, implies a factor of $2$ decrease in the growth rate.)  In
plasma traveling at the typical outflow velocity of
$v_{\text{out}}\approx 1$ the lengthscale associated with one
$e$-folding of the instability is $\ell \approx v_{\text{out}}/\gamma
= 5$.  A current layer of half-length $\Delta = 60$ can then fit
$\Delta/\ell \approx 12$ growth periods within it, more than enough to
amplify any initial fluctuations to non-linear amplitudes.

A fundamental question is whether the dynamics of the Weibel
instability scale in a manner such that the resultant length of the
current layer can produce reconnection rates insensitive to the size
of the computational domain. A complete model should incorporate the
development of the temperature anisotropy in the outflow from the
X-point by using the full dispersion relation of equation \ref{disp}.
We consider instead a cruder model in which we ignore the
anisotropy threshold. That is, we take the plasma near the X-point to
be cold so that the Weibel mode is unstable as soon as the anisotropy
develops downstream. Strictly speaking this assumption does not apply
near the X-point and will lead to an overstatement of the mode's
growth rate.  But since the principal effects of the instability
manifest themselves in the downstream region, the key result derived
from this simpler approach --- the independence of the reconnection
rate on the system size --- should remain valid in a more complete
theory.

In its most primitive form, neglecting onset thresholds, the growth
rate of the Weibel is given by $\gamma = v_{\text{th},x}/d_e$. In
Figure \ref{texx} we show that the effective electron thermal speed in
the $x$-direction is related to the relative drift between the
basically stationary upstream plasma crossing into the outflow jet and
the high speed particles accelerated from the X-point. Hence
$v_{\text{th},x}$ is simply one-half the drift speed of those fast
particles, which implies $\gamma= v_x/2d_e$. As discussed previously,
the amplitude of the turbulent fluctuations is controlled by the
convective amplification of the initial electromagnetic
perturbations. The number of $e$-foldings between the X-point and an
position $x$ downstream is therefore given by
\begin{equation}
\Gamma=\int^x_0\gamma\,\frac{dx}{v_x}.
\end{equation}
But the growth rate $\gamma$ is itself proportional to $v_x$, so $v_x$
drops out of the calculation and
\begin{equation}
\Gamma=x/2d_e.
\end{equation}
If we assume that $\Gamma_0$ is the number of $e$-foldings required to
amplify initial fluctuations to the values of order unity required to
broaden the current layer, we obtain the length of the current layer
$\Delta=2\Gamma_0d_e$.  The resulting rate of reconnection can then be
calculated from equation \ref{cont} (ignoring density changes and any
variations in the current layer width) to be
\begin{equation}
v_{\text{in}} \sim v_A/\Gamma_0.
\label{rate}
\end{equation}
The rate of reconnection resulting from the broadening of the current
layer by Weibel generated turbulence should therefore remain fast,
independent of system size, and independent of kinetic scales, as we
see in our simulations

\section{Discussion}\label{discussion}

We have shown that reconnection in a pair plasma remains fast despite
the absence of the Hall term.  For small systems, up to $100d_e$ along
the outflow direction, the length of the current layer that develops
during reconnection scales with the size of the system, as expected
based on the classical Sweet-Parker model with no dispersive waves.
When the domain size exceeds $\approx 200d_e$ a Weibel-like instability
develops that forces the development of a turbulent outflow jet and
limits the length of the narrow current layer, allowing reconnection
to remain fast.  Note that, despite being big enough to show the
transition to system-size independent reconnection, our simulation
boxes remain small in astrophysical terms.  A scale of $L=10^3$
inertial lengths, modestly larger than our biggest boxes, corresponds
to
\begin{equation}
L \sim 5\times10^3 \left(\frac{1\medspace
\text{cm}^{-3}}{n_e}\right)^{1/2} \text{km}
\end{equation}
The length $\Delta \approx 70\, d_e$ and width $\delta \approx 5\,
d_e$ of the current layer are even smaller still.  The wide separation
of scale between the system size and the scale where the important
reconnection physics occurs is a well-known phenomenon in
magnetospheric physics where in situ satellite measurements have
studied the inner scales near the X-point \citep{mozer02a}.

\begin{figure}
\plotone{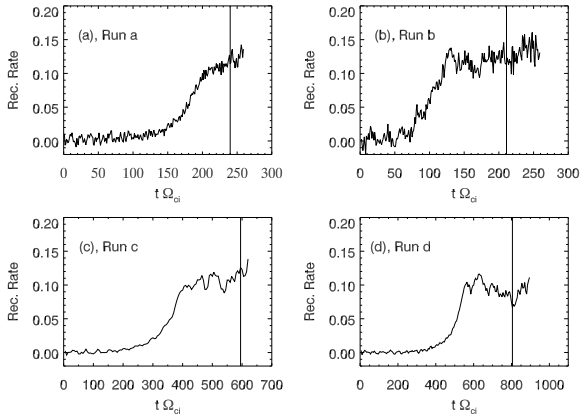}
\caption{Reconnection rate versus time for the runs listed in 
Table \ref{tbl-1}.  The vertical lines indicate the times of the
snapshots shown in Figure \ref{compare}.  Note the different scales on
the horizontal axes.  Differences in the strength of the initial
perturbation and the initial width of the current layer account for
the differences in ramp-up times between the panels.\label{rec_rate}}
\end{figure}

\begin{figure}
\plotone{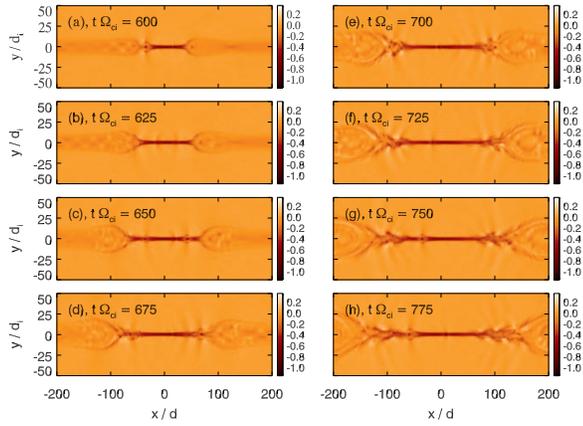}
\caption{Out-of-plane electron velocity at $\delta t = 25$ intervals
beginning at $t = 600$ for run $d$.\label{time}}
\end{figure}

Although fast pair reconnection has been noted in previous work, the
underlying cause was attributed to different mechanisms.  In
particular \cite{ daughton07a} suggest that the generation and
ejection of magnetic islands play an important role by continually
disrupting the growth of a current layer that would otherwise scale
with the system size.  Magnetic islands form in our simulations, but
appear to play little or no role in maintaining the length of the
current layer.  Figure \ref{time} shows snapshots at intervals of
$\Omega_{ci}\delta t = 25$ of the out-of-plane electron velocity for
run $d$, beginning at $\Omega_{ci}t = 600$.  During this period the
reconnection rate remains steady and fast.  In the first four panels
the instability described in section \ref{instability} does not have
enough room to grow and the current layer lengthens, in a process
analogous to what occurs in our smaller simulation domains.  In the
second column the current layer grows long enough for the Weibel-like
instability to saturate and stop further lengthening of the narrow
portion of the current layer.  Although several islands form and
travel downstream during this time, they do not appear to have any
significant effect on the length of the current layer or on the
system's reconnection rate (see Figure \ref{rec_rate}).  However,
several factors may play a role in island generation in numerical
simulations including the initial width of the current sheet, the
strength of the initial perturbation, the number of simulated
particles, and the size of the grid.  The relative importance of these
various effects, and what they imply about island generation in real
plasmas, remains unknown.

A surprise of the simulations of pair reconnection shown here and in
earlier work is that, just as in the electron-proton system, the
inflow velocity is $\mathcal{O}(0.1c_A)$.  This agreement is
unexpected since different mechanisms are clearly at play in the two
cases --- while there is some turbulence in the outflow jets of
electron-proton reconnection, the amplitude is far smaller than in the
pair system.  However, since the growth of the Weibel instability
appears to be crucial for maintaining fast reconnection in our pair
plasma simulations, any suppression of the growth rate would serve to
slow the process.  One possible mechanism for doing so is the addition
of a constant guide field component, $B_z$.  Such a field would tend
to isotropize the temperatures in the current layer ($T_x
\approx T_y$) simply by keeping the particles on Larmor orbits and
either substantially decrease or stabilize the growth of the
instability. On the other hand, we also cannot rule out the
possibility that another instability would develop when a guide field
is present. It seems clear that any elongated current layer will
ultimately develop strong pressure anisotropies, and in the presence
of a guide field the parallel pressure $p_\|$ will likely become much
larger than that perpendicular to the magnetic field $p_\bot$. For
randomly oriented magnetic fields most reconnection is guide field
reconnection, so the anti-parallel geometry we consider here is in
some sense a special case.  But, if guide field reconnection is slow
due to the suppression of the anisotropy instabilities, anti-parallel
reconnection, even if rare, may dominate in pair plasmas.

\begin{figure}
\plotone{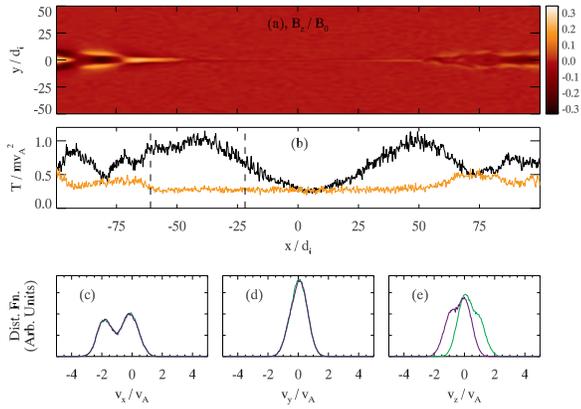}
\caption{Panel (a): $B_z$ at $t=450$, soon after the instability
develops, for run $c$. Panel (b): Two components, $T_{xx}$ and
$T_{yy}$, of the positron temperature tensor along the line $y = -1$.
Panels (c)--(e): The velocity distributions taken between the dashed
lines in panel (b) and $-1.5<y<-0.5$.  Electrons are green, positrons
blue.  In panels (c) and (d) the two curves are nearly identical.\label{texx}}
\end{figure}

Even during anti-parallel reconnection the Weibel instability will be
suppressed for large enough asymptotic plasma temperatures.  As can be
seen in Figure \ref{texx} a large $T$ can minimize the effects of the
extra heating provided by the beam of outgoing plasma in the current
layer.  The critical temperature can be roughly estimated as $T_{crit}
\approx m v_d^2 /2$ where $v_d\sim c_A$ is the drift velocity of the
outflowing electrons.  Thus for asymptotic temperatures $T > T_{crit}
\approx mc_A^2$ or $\beta >1$, the instability should be
stabilized. Further simulations in this regime are being performed.

\acknowledgements

After submitting this paper we became aware that Zenitani and Hesse
also concluded that the Weibel instabilty plays a similar role during
reconnection in very relativistic pair plasmas \citep{zenitani08a}.

\clearpage

\clearpage

\clearpage

\clearpage

\clearpage

\clearpage

\clearpage

\begin{table}
\begin{center}
\caption{Simulation parameters.\label{tbl-1}}
\begin{tabular}{cccc}
\tableline\tableline
Run Label & Domain Size & Gridpoints & $w_0$ \\
\tableline
a & $100 \times 50$ & $512 \times 256$ & 2\\
b & $200 \times 100$ & $1024 \times 512$ & 2\\
c & $400 \times 200$ & $2048 \times 1024$ & 4\\
d & $800 \times 200$ & $4096 \times 1024$ & 4\\
\tableline
\end{tabular}
\tablecomments{$w_0$ is the initial half-width of the current sheet.}

\end{center}
\end{table}

\begin{table}
\begin{center}
\caption{Simulation results.\label{tbl-2}}
\begin{tabular}{cccccccc}
\tableline\tableline
Run Label & $n_{\text{in}}$ & $n_{\text{out}}$ & $2\delta$ & $2\Delta$ 
& $v_{\text{out}}$ & $v_{\text{in,meas}}$ & $v_{\text{in,calc}}$ \\
\tableline
a & 0.16 & 0.27  & 4.0 & 35  & 0.5  & 0.13 & 0.10 \\
b & 0.12 & 0.32 & 4.0 & 80 & 0.8 & 0.15 & 0.11 \\
c & 0.13 & 0.33 & 4.5 & 120 & 1.3 & 0.16 & 0.12 \\
d & 0.13 & 0.30 & 5.0 & 135 & 1.3 & 0.13 & 0.11 \\
\tableline
\end{tabular}
\tablecomments{2$\delta$ and 2$\Delta$ denote the smaller and larger
spatial extents, respectively, of the current
sheet. $v_{\text{in,meas}}$ and $v_{\text{in,calc}}$ are the measured
and calculated (from equation \ref{cont}) velocities of plasma into
the X-point.}

\end{center}
\end{table}


\begin{thebibliography}{27}
\providecommand{\natexlab}[1]{#1}
\expandafter\ifx\csname urlstyle\endcsname\relax
  \providecommand{\doi}[1]{doi:\discretionary{}{}{}#1}\else
  \providecommand{\doi}{doi:\discretionary{}{}{}\begingroup
  \urlstyle{rm}\Url}\fi

\bibitem[{\textit{Bessho and Bhattacharjee}(2005)}]{bessho05a}
Bessho, N., and A.~Bhattacharjee (2005), Collisionless reconnection in an
  electron-positron plasma, \textit{Phys. Rev. Lett.}, \textit{95}(24), 245001,
  \doi{10.1103/PhysRevLett.95.245001}.

\bibitem[{\textit{Bessho and Bhattacharjee}(2007)}]{bessho07a}
Bessho, N., and A.~Bhattacharjee (2007), Fast collisionless reconnection in
  electron-positron plasmas, \textit{Phys. Plasmas}, \textit{14}(05), 056503,
  \doi{10.1063/1.2714020}.

\bibitem[{\textit{Birn et~al.}(2001)}]{birn01a}
Birn, J., et~al. (2001), Geospace {E}nvironmental {M}odeling ({GEM}) magnetic
  reconnection challenge, \textit{J. Geophys. Res.}, \textit{106}(A3),
  3715--3719.

\bibitem[{\textit{Biskamp and Schwarz}(2001)}]{biskamp01a}
Biskamp, D., and E.~Schwarz (2001), Localization, the clue to fast magnetic
  reconnection, \textit{Phys. Plasmas}, \textit{8}(11), 4729--4731,
  \doi{10.1063/1.1412600}.

\bibitem[{\textit{Coroniti}(1990)}]{coroniti90a}
Coroniti, F.~V. (1990), Magnetically striped relativistic magnetohydrodynamic
  winds: The {C}rab nebula revisited, \textit{Astrophys. J.}, \textit{349},
  538--545.

\bibitem[{\textit{Daughton and Karimabadi}(2007)}]{daughton07a}
Daughton, W., and H.~Karimabadi (2007), Collisionless magnetic reconnection in
  large-scale electron-positron plasmas, \textit{Phys. Plasmas}, \textit{14},
  072303, \doi{10.1063/1.2749494}.

\bibitem[{\textit{Drenkhahn and Spruit}(2002)}]{drenkhahn02a}
Drenkhahn, G., and H.~C. Spruit (2002), Efficient acceleration and readiation
  in {P}oynting flux powered {GRB} outflows, \textit{Astronomy \&
  Astrophysics}, \textit{391}, 1141--1153, \doi{10.1051/0004-6361:20020839}.

\bibitem[{\textit{Eardley and Lightman}(1975)}]{eardley75a}
Eardley, D.~M., and A.~P. Lightman (1975), Magnetic viscosity in relativistic
  accretion disks, \textit{Astrophys. J.}, \textit{200}, 187--203.

\bibitem[{\textit{Fujimoto}(2006)}]{fujimoto06a}
Fujimoto, K. (2006), Time evolution of the electron diffusion region and the
  reconnection rate in fully kinetic and large system, \textit{Phys. Plasmas},
  \textit{13}(7), 072904, \doi{10.1063/1.2220534}.

\bibitem[{\textit{Harris}(1962)}]{harris62a}
Harris, E.~G. (1962), On a plasma sheet separating regions of oppositely
  directed magnetic field, \textit{Nuovo Cim.}, \textit{23}, 115.

\bibitem[{\textit{Hesse et~al.}(2001)\textit{Hesse, Kuznetsova, and
  Birn}}]{hesse01a}
Hesse, M., M.~Kuznetsova, and J.~Birn (2001), Particle-in-cell simulations of
  three-dimensional collisionless magnetic reconnection, \textit{J. Geophys.
  Res.}, \textit{106}(A12), 29,831--29,842.

\bibitem[{\textit{Jaroschek et~al.}(2004)\textit{Jaroschek, Treumann, Lesch,
  and Scholer}}]{jaroschek04a}
Jaroschek, C.~H., R.~A. Treumann, H.~Lesch, and M.~Scholer (2004), Fast
  reconnection in relativistic pair plasmas: Analysis of particle acceleration
  in self-consistent full particle simulations, \textit{Phys. Plasmas},
  \textit{11}(3), 1151--1163, \doi{10.1063/1.1644814}.

\bibitem[{\textit{Krall and Trivelpiece}(1986)}]{krall86b}
Krall, N.~A., and A.~W. Trivelpiece (1986), \textit{Principles of Plasma
  Physics}, chap.~9, pp. 483--494, San Francisco Press, Inc.

\bibitem[{\textit{Masuda et~al.}(1994)\textit{Masuda, Kosugi, H, Tsuneta, and
  Ogawara}}]{masuda94a}
Masuda, S., T.~Kosugi, .~H. H, S.~Tsuneta, and Y.~Ogawara (1994), A loop-top
  hard x-ray source in a compact solar flare as evidence for magnetic
  reconnection, \textit{Nature}, \textit{371}, 495--497,
  \doi{10.1038/371495a0}.

\bibitem[{\textit{Michel}(1994)}]{michel94a}
Michel, F.~C. (1994), Magnetic structure of pulsar winds, \textit{Astrophys.
  J.}, \textit{431}, 397--401.

\bibitem[{\textit{Miller and Stone}(1997)}]{miller97a}
Miller, K.~A., and J.~M. Stone (1997), Magnetohydrodynamic simulations of
  stellar magnetosphere-accretion disk interaction, \textit{Astrophys. J.},
  \textit{489}, 890--902.

\bibitem[{\textit{Mozer et~al.}(2002)\textit{Mozer, Bale, and Phan}}]{mozer02a}
Mozer, F.~S., S.~D. Bale, and T.~D. Phan (2002), Evidence of diffusion regions
  at a subsolar magnetopause crossing, \textit{Phys. Rev. Lett.}, \textit{89},
  \doi{10.1103/PhysRevLett89.015002}.

\bibitem[{\textit{Parker}(1957)}]{parker57b}
Parker, E.~N. (1957), Sweet's mechanism for merging magnetic fields in
  conducting fluids, \textit{J. Geophys. Res.}, \textit{62}(4), 509--520.

\bibitem[{\textit{Petschek}(1964)}]{petschek64a}
Petschek, H.~E. (1964), Magnetic field annihilation, in \textit{Proc. AAS-NASA
  Symp. Phys. Solar Flares}, \textit{NASA-SP}, vol.~50, pp. 425--439.

\bibitem[{\textit{Phan et~al.}(2006)}]{phan06a}
Phan, T.~D., et~al. (2006), A magnetic reconnection {X}-line extending more
  than 390 {E}arth radii in the solar wind, \textit{Nature}, \textit{439},
  175--178, \doi{10.1038/nature04393}.

\bibitem[{\textit{Rogers et~al.}(2001)\textit{Rogers, Denton, Drake, and
  Shay}}]{rogers01a}
Rogers, B.~N., R.~E. Denton, J.~F. Drake, and M.~A. Shay (2001), The role of
  dispersive waves in collisionless magnetic reconnection, \textit{Phys. Rev.
  Lett.}, \textit{87}(19), 195,004.

\bibitem[{\textit{Shay et~al.}(1999)\textit{Shay, Drake, Rogers, and
  Denton}}]{shay99a}
Shay, M.~A., J.~F. Drake, B.~N. Rogers, and R.~E. Denton (1999), The scaling of
  collisionless, magnetic reconnection for large systems, \textit{Geophys. Res.
  Lett.}, \textit{26}(14), 2163--2166.

\bibitem[{\textit{Shay et~al.}(2007)\textit{Shay, Drake, and
  Swisdak}}]{shay07a}
Shay, M.~A., J.~F. Drake, and M.~Swisdak (2007), Two-scale structure of the
  electron dissipation region during collisionless magnetic reconnection,
  \textit{Phys. Rev. Lett.}, \textit{99}, 155002,
  \doi{10.1103/PhysRevLett.99.155002}.

\bibitem[{\textit{Sweet}(1958)}]{sweet58a}
Sweet, P.~A. (1958), \textit{Electromagnetic Phenomena in Cosmical Physics}, p.
  123, Cambridge University Press, New York.

\bibitem[{\textit{Zeiler et~al.}(2002)\textit{Zeiler, Biskamp, Drake, Rogers,
  Shay, and Scholer}}]{zeiler02a}
Zeiler, A., D.~Biskamp, J.~F. Drake, B.~N. Rogers, M.~A. Shay, and M.~Scholer
  (2002), Three-dimensional particle simulations of collisionless magnetic
  reconnection, \textit{J. Geophys. Res.}, \textit{107}(A9), 1230,
  \doi{10.1029/2001JA000287}.

\bibitem[{\textit{Zenitani and Hoshino}(2001)}]{zenitani01a}
Zenitani, S., and M.~Hoshino (2001), The generation of nonthermal particles in
  the relativistic magnetic reconnection of pair plasmas, \textit{Astrophys. J.
  Lett.}, \textit{562}, L63--L66.

\bibitem[{\textit{Zenitani and Hesse}(2008)}]{zenitani08a}
Zenitani, S., and M.~Hesse (2008), The role of the Weibel instability
  at the reconnection jet front in relativistic pair plasma reconnection,
  \textit{Phys. Plasmas}, in press.

\end{thebibliography}
\end{document}